\documentclass[11pt]{article}
\usepackage{graphicx,pstricks}
\usepackage{booktabs}
\usepackage{amsmath, amsthm, amssymb}

\title{\LARGE \bf The GHSZ argument: a gedankenexperiment requiring more denken}

\author{Frank Lad, \ \ \ email:   \ frank.lad@canterbury.ac.nz \\
{\small University of Canterbury, Department of Mathematics and Statistics}}

\date{17 December, 2020}

\begin{document}

\maketitle

\begin{abstract}
%\noindent I reassess the influential gedankenexperiment of Greenberger, Horne, Shimony and Zeilinger after twenty-five years, finding their claim
%to discovery of an inconsistency inherent in the principle of local realism to arise from
%a fundamental error of simple logic.  They themselves introduced the contradiction they had purported to discover in high dimensional
%formulations of local realism.
%Even more extensively, their analysis of a 4-ply experiment involving electron spins engenders a miasma of errors
%that derive from incomplete notation and quite casual thinking.  Ignoring quantum entanglement among the four particles is the source of their
%mistaken and misleading conclusions.  In tandem with the error now recognized in the supposed defiance of Bell's inequality by quantum
%probabilities, my reassessment of their work should motivate a reevaluation of the current consensus
%outlook regarding the principle of local realism and the proposition of hidden variables.
%Careful considerations show how to generate the consistent EPR model of four-dimensional spin for which GHSZ searched.\\

\noindent I reassess the gedankenexperiment of Greenberger, Horne, Shimony and Zeilinger after twenty-five years, finding their influential claim
to discovery of an inconsistency inherent in  high dimensional
formulations of local realism to arise from
a fundamental error of logic.  They manage this by presuming contradictory premises:   that a specific linear combination of four angles 
involved in their proposed parallel experiments
  on two pairs of electrons equals both $\pi$ and $0$ at the same time.  Ignoring this while presuming the contradictory implications of these two  
  conditions, they introduce the contradiction themselves.  The notation they use in their ``derivation'' is not sufficiently ornate  to represent  
  the entanglement in the double electron spin pair problem they design, confounding their error.  The situation they propose actually
  motivates only an understanding of the full array of symmetries involved in their problem.  In tandem with the error now recognized in the 
  supposed defiance of Bell's inequality by quantum
probabilities, my reassessment of their work should motivate a reevaluation of the current consensus
outlook regarding the principle of local realism and the proposition of hidden variables.

%\noindent  GHSZ (Greenberger, Horne, Shimony, and Zeilinger, 1991) "find" their contradiction because they insert it into their analysis themselves by 
 % presuming contradictory premises:   that the linear combination of the angles $\phi_1+\phi_2-phi_3-phi_4$ involved in their proposed parallel experiments 
 % on two pairs of photons equals both $\pi$ and $0$ at the same time.  Ignoring this while presuming the contradictory implications of these two 
  %conditions, they introduce the contradiction themselves.  The notation they use in their ``derivation'' is not sufficiently ornate  to represent 
  %the entanglement in the double electron spin pair problem they design, confounding their error of neglect.  The situation they propose actually 
  %motivates only an understanding of the full array of symmetries involved in their problem.  

\end{abstract}


\section {Prelude}

When I first began to disseminate the results of my analysis
that refute the supposed defiance of Bell's inequality by quantum probabilities (Lad, 2020a), I was immediately challenged by suspicious and offended physicists.  Considered to be incomprehensible and
confused, the focus of my analysis was  relegated as being completely out of date with subsequent developments.
Not only was
Aspect's experimental setup deemed to be crude and to have been superceded,
but it was proposed to me that a definitive analysis
by Greenberger, Horne, Shimony and Zeilinger (1990) had determined that the premises of the EPR argument entailing
``perfect correlation, reality, locality, and completeness'' could not
even be modeled consistently in quantum problems involving more than two dimensions. GHSZ were said to have
 expounded a version of Bell's theorem without involving inequalities.  Aware of my evident naivet\'{e}
regarding matters I had long avoided in my researches, and keen to learn what this might be about, I eagerly dove
right in!\\

Upon reading, I was stunned to learn that their claims, already studied and honoured within the physics
community for some
twenty-five years, are based on an argument that defies a simple rule of deductive logic,
and is entrained in cryptic notational
sleight of hand which has misled the leading researchers of the scientific community. Their article will have been
 read by surely more than five thousand physicists and students during this time.  It is with some embarassment
that I shall explain these miscues in this article.  I invite you to please continue reading, with an open mind, if only
out of curiosity.  \\

Any reader prepared to do so is advised to refresh yourself with the GHSZ
argument in their own words before continuing.  However, to make this presentation
self-contained, their argument will be developed here faithfully until the point that it hits the fan, using their notation precisely until it will become necessary to refine and embellish it. I use their numbering system for equations, beginning with (7), the first numbered equation of their Section III
which is the sole subject matter of this exposition. When their equations are repeated using a refined notation, their same equation numbers are merely adorned with a superscripted star$^*$.  Lower case Roman numerals are employed 
 to designate the few new equations that are introduced.  I mention in passing only that the persuasive exposition of Bell's original Theorem in the Section II of
the GHSZ article continues to embed the error which Bell and subsequent advocates of the supposed violation have missed. \\

My deferential attitude to the reader's review of the original paper, and the care with which I shall now assess it, are appropriate to the influential role the article has played in motivating the current general understanding of the quantum scale defiance of Bell's inequality within the physics community.  Of course the inequality itself has provoked extensive analysis and a myriad of commentaries.  Notable among them are the comprehensive review by Brunner et al (2014) and wide-ranging discussions in the texts of Greenstein and Zajonc (2006) and of Jaeger (2009).  Within this context, the provocative claims of GHSZ to a contradiction involved in the formulation of theories of local realism are considered to be definitive.  They form part of a stimulating literature that stems from the original challenging work of von Neumann (1932, 1955) and runs through Mermin and Schack (2018).  While subscription to various arguments has varied over time, the concluding assessment of GHSZ by Jaeger (2009, p. 44-45) is supported by virtually all researchers whose work is referenced in these reviews: 
	\begin{quote} Remarkably,
		a decade after Aspect’s tests of the CHSH inequality, it was shown by
		Greenberger, Horne, Shimony, and Zeilinger (GHSZ) that the premises of the
		Einstein–Podolsky–Rosen paper become inconsistent when applied to systems
		possessing three or more subsystems, even for the cases involving such perfect
		correlations [194].
		The GHSZ demonstration shows that the incompatibility of the EPR assumptions
		with quantum mechanics is stronger than that indicated by the
		violation of the Bell and CHSH inequalities, in that in the case of a pair of
		two-level systems there is no internal contradiction at the level of perfect correlations.
		Indeed, Bell produced an explicit model for the case of a pair of
		spin-1/2 particles demonstrating the consistency of the EPR conditions with
		the perfect correlations predicted by quantum mechanics [23]. Furthermore,
		the contradiction between quantum mechanical predictions and the Bell and
		CHSH inequalities are expressions violated only by statistical predictions of
		quantum mechanics, rather than by individual events.
		In the lead up to the exceptionally clear exposition of GHSZ, Greenberger,
		Horne, and Zeilinger (GHZ) demonstrated the inconsistency in a new way in
		systems consisting of three or more correlated spin-1/2 particles [195]. Because
		this showed that the incompatibility of quantum mechanics with the EPR
		assumptions arises at the level of perfect correlations rather than statistical
		predictions and did not require the use of an inequality, these results are
		often referred to as “Bell’s theorem without inequalities.” 
		\end{quote}
In a word, I am fully aware of the momentous content of the analysis I shall present here, and the extent of the 
reconsiderations it should require.  While my presentation shall be lively, it is not with bravado but with honest concern that I suggest we have been deeply engaged in a serious mistake for some fifty years since the advent of Bell's first  publications on the matter, a mistake he himself had suspected but could not identify. If I am in error it will need to be displayed by reasoning rather than by reverence for vaunted old understandings. We all make mistakes. Let's get into it. \\

\section {The physical setup of the 4-D GHSZ experiment}

The context of the GHSZ experiment does not involve the behaviour of photons, but rather of four electrons that are
propelled toward Stern-Gerlach analyzing magnets so to determine the directions of their orbital spins.  The physical
setup is caricatured by their Figure 2 which is reproduced here.

\begin{figure}[!h]
\begin{center}
\includegraphics[width=1\linewidth]{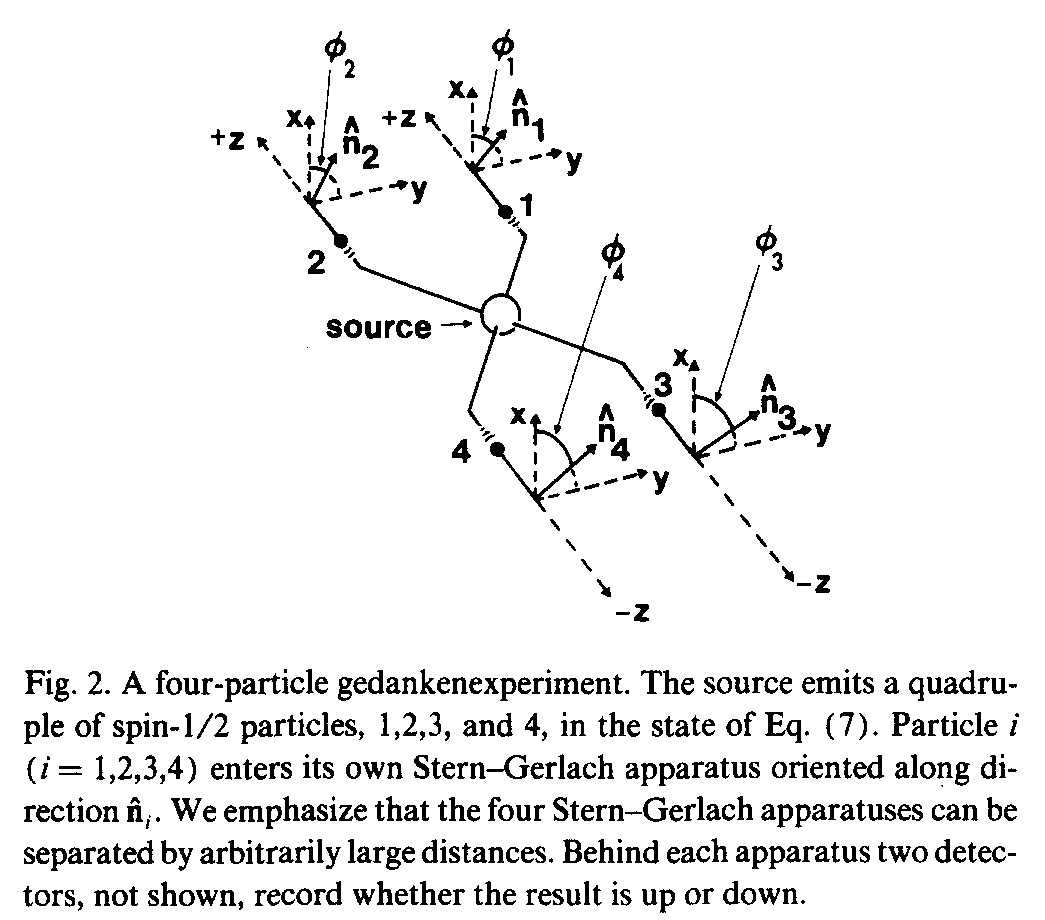}
%\includegraphics[scale=0.6]{figure2forfrank.png}
%\caption{Directional vectors of the polarization angle settings at the observation
%stations $A$ and $B$, viewed in a common axis orientation.  The specific relative angle
%size settings displayed are the extreme violation values, a feature to be discussed.}
\caption{The original schema of the GHSZ article, reproduced with permission from
{\it American Journal of Physics},  {\bf 58}(12), page 1134}
\label{fig:xtrmviolationanglesFig2}
\end{center}
\end{figure}

This schema represents a physical process in which one pair of electrons (referred to as ``spin-1/2 particles'')
are transmitted in spatially separated beams in the direction ${\bf z^+}$ toward Stern-Gerlach analyzers at stations
$1$ and $2$ while another pair are transmitted far away in the opposite direction ${\bf z^-}$ toward similarly separated
analyzers at stations $3$ and $4$.  The electrons are said to be in a superposition of two states vis-\`a-vis their spins,
implying that they
might be observed as either ``up'' or ``down'' when detected at any analyzer. As transmitted, they are
 considered to be in both and neither states, superposed.  The observation ``up'' is designated
by a recording of $+1$, while ``down'' is recorded as $-1$.  Now the Stern-Gerlach magnets can be positioned in various
directions relative to that in which the incoming particles arrive. These directional vectors are denoted in the
Figure as ${\hat{\bf n}_1, \hat{\bf n}_2, \hat{\bf n}_3}$, and $\hat{\bf n}_4$.
 In the $(x,y)$ plane perpendicular to any
${\bf z}$ direction, the Figure denotes by $\phi$ the angle relative to the $x-$dimension in which any one of them
is directed.  The four angle sizes $\phi_1$, $\phi_2$, $\phi_3$, and $\phi_4$ can be variable, but they are
fixed at specific values for any run of the experiment.  Of course the $(x,y)$ planes of these detectors could also be
twisted toward the direction of their ${\bf z}$ axes as well.  Such can be accounted for in the theory according
to the GHSZ equation $(8)$ in which these angles are denoted by $\theta_i$.  However, they are not depicted in the Figure, for
throughout this discussion they are considered to be fixed at $\theta_i = \pi/2 = 90^{\circ}$,
each of the four $(x,y)$ planes being perpendicular to the direction of ${\bf z}$.  This reduces their general
equation $(8)$ to $(9)$ which we shall examine shortly.\\

%${\bf \left[ \, } {\bf \right] \,}
The 4-component electron propagation
mechanism, which we shall not discuss here, ensures the spin state of the four-plex of electrons is superposed
according to the prescription  \vspace{.2cm} \\
\noindent \hspace*{2cm}$|{\bf \Psi}\rangle  \ \ = \ \
(1/\sqrt{2})\;  \left[ \; |+\rangle_1\:|+\rangle_2\:|-\rangle_3\:|-\rangle_4 \:\ + \ \:|-\rangle_1|\:|-\rangle_2|\:|+\rangle_3|\:|+\rangle_4\;\right] \ . \ \ \ \ \ \ \ \ \ \ \ \ \ \ \vspace{.2cm}\ (7)$  \\
If you are not au fait with the mathematical apparatus of quantum theory, please do not be put off by
this notation, and read on.  This is merely a concise notation for representing the structure of the
experimental situation
we have explained in the previous paragraph.  Trust yourself. You are ready to go.

%  Here is an explanatory sentence from Wikipedia regarding electron spin:  "The motion of an electric charge producing a magnetic field is an essential concept in understanding magnetism. The magnetic moment of an atom can be the result of the electron's spin, which is the electron orbital motion and a change in the orbital motion of the electrons caused by an applied magnetic field."
\subsection{Identifying the prescriptions of quantum theory}
When detections are made of the electrons' spins at stations $1, 2, 3$, and $4$, these are designated by the
variable names $A, B, C$, and $D$, respectively.  Each of these values might arise as $-1$ or $+1$ according to
the direction of the corresponding spin observation.  Now what would be the values of these observation values in
any run of the experiment?  The theory of quantum mechanics does not specify exactly what the results would be
in any run of such experiments, even though each run in a sequence might be set up carefully in exactly the same way.
Quantum theory yields only a probability distribution for the four measurement results from any run.
Now just as our investigation of the
polarized photon pair detection in the Aspect/Bell experiment
found the crucial feature of the observations was the value of the {\it product} of the
detection results, the crucial part of the QM specifications of probabilities for observation values of the four
electron spins in a run revolves about their product:  $ABCD$.\\

The product of the spin measurements in any run depends stochastically on the four angle settings in the
design of that run.  GHSZ derive the mathematical expectation of this spin-product in Appendix $F$ of
their article for the general case in which the angles $\theta_i$ are variable.
In the article itself they simplify this expectation to the special case in which each
of the $(x,y)$ planes is perpendicular to its relevant $z$ axis.
%yielding  $E^{{\bf \psi}}[ABCD(\phi_1,\phi_2,\phi_3,\phi_4)] \ = \ - cos(\phi_1 + \phi_2 - \phi_3 - \phi_4)$.
Using cryptic notation, the spin-product expectation is reported for this setup in a numbered equation  as\vspace{.2cm}\\
\hspace*{4cm} $E^{{\bf \psi}}(\hat{{\bf n}}_1, \hat{{\bf n}}_2, \hat{{\bf n}}_3, \hat{{\bf n}}_4) \ = \ - cos(\phi_1 + \phi_2 - \phi_3 - \phi_4)\ \ . \hspace{2cm} \vspace{.2cm}\ \ \ \ \ \ (9)$  \\
\noindent They describe this equation as ``the expectation of the product of the outcomes when the orientations
are as indicated''.
To be sure, this notation has a sensible motivation.  For it pertains to
an expectation of a function (the product) of spin observations on a system of particles designed to reside
in the superposed state designated by $|{\bf \Psi}\rangle$ in which the Stern-Gerlach analyzers are positioned
at angles corresponding to the directional vectors ${\hat{\bf n}_1, \hat{\bf n}_2, \hat{\bf n}_3}$, and $\hat{\bf n}_4$.
These directional vectors determine the angles $\phi_1, \phi_2, \phi_3$, and $\phi_4$ whose linear combination is the
argument of the cosine function in $(9)$ that specifies the expectation value.  Well, I do not think I am
being picayune, and I hope I am not, in highlighting that this expectation
$E^{{\bf \psi}}(\hat{{\bf n}}_1, \hat{{\bf n}}_2, \hat{{\bf n}}_3, \hat{{\bf n}}_4)$ is {\it not} the
expectation of a vector of four directional vectors (which it ostensibly is),
but rather it is the expectation of the product of four spin observation values in an experiment at which the
analyzing magnets are aligned with these directions. It would be depicted more clearly
as identifying $E^{{\bf \psi}}[ABCD(\phi_1,\phi_2,\phi_3,\phi_4)]$.
 While understandable, the notation they use serves
to obscure the sensible analysis of the situation, as we shall now discover. \\

Of particular interest to the GHSZ argument are the cases of perfect correlation, which they designate by
the conditioned equations\vspace*{.2cm}\\
\hspace*{5cm}If\ \ \ \ $\phi_1 + \phi_2 - \phi_3 - \phi_4 \ = \ 0 $\vspace*{.2cm}\\
\hspace*{5cm}then \ \ \ $E^{{\bf \psi}}(\hat{{\bf n}}_1, \hat{{\bf n}}_2, \hat{{\bf n}}_3, \hat{{\bf n}}_4) \ = \ -1\,, \ \ \ \ \ \vspace{.25cm}(10a)$\\
\hspace*{5cm}If \ \ \ $\phi_1 + \phi_2 - \phi_3 - \phi_4 \ = \ \pi $\vspace*{.2cm}\\
\hspace*{5cm}then \ \ \ $E^{{\bf \psi}}(\hat{{\bf n}}_1, \hat{{\bf n}}_2, \hat{{\bf n}}_3, \hat{{\bf n}}_4) \ = \ +1\,. \ \ \ \ \ \vspace{.2cm}(10b)$\\
The conditions are printed exactly like that, in four lines of which the second and fourth lines conclude with the
equation number designations $(10a)$ and $(10b)$.  This unfortunate feature of the double column print style
of their publication will be found relevant to their mistaken argument.  A second unfortunate feature of their
considerations is that they did not designate explicitly in notation the statement that they described in words:
that the single observed product $ABCD$ of the spin recordings depends (stochastically) on the {\it four} orientations of the detection magnets,
$(\phi_1,\phi_2,\phi_3,\phi_4)$.  Mentioning for now only that these two misfortunes will return to haunt us, we shall
continue with the GHSZ argument.\\

Interest in the conditions of $(10a)$ and $(10b)$ arises from the fact that they support the supposition of Einstein, Podolsky, and
Rosen (EPR, 1938) of perfect correlation,
a condition specified by the derivations of quantum theory.  The other three suppositions of EPR are noted
to concern matters extraneous to quantum theory, though they are recognised as plausible and in agreement with
principles of classical physics.\\

GHSZ were pleased that this experimental design might instantiate the requirements of Bell's inequality
in a four-dimensional problem.  However, they proceeded no further in pursuing details of this matter, because they first pursued
an argument that the structure developed to this point embeds a contradiction among the four premises of EPR,
these being (i) perfect correlation, (ii) reality, (iii) locality, and (iv) completeness.
Let's begin to follow their line of argument.

\subsection{Pursuing the contradiction discovered by GHSZ}

They begin by restating the conditional statements $(10a)$ and $(10b)$ but now using functional observation notations
$A(^.),\ B(^.),\ C(^.)$, and $D(^.)$ in
stating the conclusions of the ``If'' clauses.  They write \vspace*{.2cm}\\
\hspace*{4cm}If \ \ \ $\phi_1 + \phi_2 - \phi_3 - \phi_4 \ = \ 0 $\vspace*{.2cm}\\
\hspace*{4cm}then \ \ \ $A_{\lambda}(\phi_1)B_{\lambda}(\phi_2)C_{\lambda}(\phi_3)D_{\lambda}(\phi_4) \ = \ -1\,, \ \ \ \ \ \vspace{.25cm}(11a)$\\
\hspace*{4cm}If \ \ \ $\phi_1 + \phi_2 - \phi_3 - \phi_4 \ = \ \pi $\vspace*{.25cm}\\
\hspace*{4cm}then \ \ \ $A_{\lambda}(\phi_1)B_{\lambda}(\phi_2)C_{\lambda}(\phi_3)D_{\lambda}(\phi_4) \ = \ +1\,. \ \ \ \ \ \vspace{.2cm}(11b)$\\
Before continuing with their development, I must make two remarks, either of which might be ignored, but both of which are of some meritorious consequence.\\

The first concerns an unspoken argument.  Notice that the lines numbered $(10a)$ and $(10b)$ make statements about
the values of {\it expectations of the products} of the spin measurements at the four observation sites, whereas lines
$(11a)$ and $(11b)$ concern the {\it products of observations themselves}. Upon consideration, this appears to be
of little consequence.  Why?
The numerical values of each multiplicand can be only either $-1$ or $+1$ by their very operational definition,
so the product of any four such observations may also equal only either $-1$ or $+1$.
Moreover, equation $(10a)$ for example quite rightly designates
 that under the condition specified in the unnumbered line just above it, the expectation must equal $-1$.  This
 implies that the probability weight attributed to the possibility that the product
  equals $+1$ must be zero!  In this condition, the product of the spins itself must equal $-1$.
 Thus, quantum theory prescribes that in this case the {\it product} of the four spin observations itself must equal $-1$, just as stated
 in equation $(11a)$.
  The very same remark would pertain to
 the replacement of $(10b)$ by $(11b)$, according to which the product of the four observations must equal $+1$
under the condition to which it pertains.\\

So changing the statement of equations $(10a,b)$ regarding expectations of a function (the product of four
measurements) to $(11a,b)$ regarding the
function values themselves appears innocuous here, because the distributions of the function value
are obviously degenerate in both cases, at
 $-1$ and $+1$ respectively.  However, we should be aware that the joint distributions of the
component multiplicands of this product have {\it not} been determined to be degenerate.  There are several arrays of
the multiplicands that could yield a product of $-1$ to instantiate $(11a)$, and several arrays that could yield a
product of $+1$ so to instantiate $(11b)$.  It is the expectation of their product which equals $-1$, and thus
their product which has a degenerate distribution.  The distribution of the four component observations in a run
remains ample.  Enough said for now, but we shall return to this recognition later.\\

The second remark is less consequential, but it is substantive relevant to the larger picture of what this
analysis of the GHSZ problem is all about.  At this point in their development, the spin observation variables
$A,\ B,\ C$, and $D$ suddenly begin to appear with the subscript $\lambda$, as in $A_{\lambda}(\phi)$.  As it turns out,
{\it every} subsequent appearance of such observation values in their article is adorned with this subscript.  Thus,
the subscript contributes nothing to the force of the relevant arguments, and I will no longer use it.  To be sure,
its presence would serve as a reminder that what we are doing here is formalising experimental matters
motivated by considerations of the EPR proposal of ``supplementary variables'' as the source of the incompleteness
of quantum theory. The vector $\lambda$ is meant to designate the unknown values of such variables.
Now this {\it is} a very important matter bearing consideration, and I am have investigated this
matter explicitly on its own in a separate article (Lad, 2020b). However for now, let us just consider ourselves
to have been reminded of this important motivation for the considerations we are pursuing, and just drop the $\lambda$
subscript, except in places where we will quote GHSZ verbatim.  For there will come a point in these considerations
at which we shall wish to use the subscript position on the spin measurements
to make explicit an important varying feature of the problem that has been ignored hitherto.\\

Let's continue then from equations $(11a)$ and $(11b)$.  I shall be following very closely the work
and even the wording of GHSZ. When exact quotations are used they will be marked.\\

Next are considered some implications of $(11a)$.   Four specific instances of this specification
are proposed, with reference to any arbitrary angle size $\phi$, as \vspace{.2cm}\\
\hspace*{4.3cm}$A(0)B(0)C(0)D(0) \ \ = \ \ -1 \vspace{.2cm} \ \ \ \ \ (12a)$\\
\hspace*{4.2cm}$A(\phi)B(0)C(\phi)D(0) \ \ = \ \ -1 \vspace{.2cm} \ \ \ \ \ (12b)$\\
\hspace*{4.2cm}$A(\phi)B(0)C(0)D(\phi) \ \ = \ \ -1 \vspace{.2cm} \ \ \ \ \ (12c)$\\
\hspace*{4cm}$A(2\phi)B(0)C(\phi)D(\phi) \ \ = \ \ -1 \vspace{.2cm}, \ \ \ \  (12d)$\\
because the four angle arguments in each of these product equations meet the condition that
$(\phi_1+\phi_2-\phi_3-\phi_4) = 0$,
which is the condition under which $(11a)$ holds.  \\

As a consequence of equalities $(12a)$ and $(12b)$ they then obtain \vspace{.2cm}\\
\hspace*{4.2cm}$A(\phi)C(\phi) \ = \ A(0)C(0) \vspace{.2cm}\ \ \ \ \ \ \ \ \ \ \ (13a)$, \\
through cancellation of $B(0)D(0)$ which appears identically on the left-hand-sides of these two
equations.  Correspondingly,  equalities $(12a)$ and $(12c)$ are seen to imply \vspace{.2cm}\\
\hspace*{4.2cm}$A(\phi)D(\phi) \ = \ A(0)D(0) \vspace{.2cm}\ \ \ \ \ \ \ \ \ \ (13b)$\ \  \\
according to similar cancellations of $B(0)C(0)$.\\

The quotients of these two results then yields for them \vspace{.2cm}\\
\hspace*{4.2cm}$C(\phi)/D(\phi) \ = \ C(0)/D(0) \vspace{.2cm}\ \ \ \ \ \ \ (14a)$\ \ , \\
which can be rewritten as \vspace{.2cm}\\
\hspace*{4.3cm}$C(\phi)\:D(\phi) \ = \ C(0)\:D(0) \vspace{.2cm}\ \ \ \ \ \ \ \ (14b)$\ \ , \\
because both $D(\phi)$ and $D(0)$ can each equal only either $+1$ or $-1$.  In either case, both of them
are equal to their inverses.\\

By substituting  $(14b)$ into $(12d)$, GHSZ then obtain the surprising result that \vspace{.2cm}\\
\hspace*{4.2cm}$A(2\phi)B(0)C(0)D(0) \ \ = \ \ -1 \vspace{.2cm} \ .\ \ \ \ (15)$\\
This, in combination with $(12a)$ which says $A(0)B(0)C(0)D(0) = -1$, yields \vspace{.2cm}\\
\hspace*{4.2cm}$A(2\phi) \ = \ A(0) \ = \ const$ for any angle $\phi$.\vspace{.2cm} \ \ \ \ \ \ \ \ \ $(16)$\\
In particular, this would imply that a measurement of $A(\pi)$ must equal $A(0)$.\\

While this result is apparently not contradictory, to a physicist well versed in
quantum theory it is surely quite troublesome.  GHSZ call it a ``preliminary result''.  Let them explain why it appears troublesome in their own words.  ``For if $A_{\lambda}(\phi)$ is intended, as EPR's program suggests, to represent an intrinsic spin quantity, then $A_{\lambda}(0)$ and $A_{\lambda}(\pi)$ would be expected to have opposite signs.''
[Be aware that an electron spin measured by analyzers positioned in two opposing directions, differing in orientation
by $\pi = 180^{\circ}$, should display
opposite directions rather than the same direction, just as two attracting magnets will repel one another if one of them is
rotated by $180^{\circ}$.]  However, this trouble is subjected to no further examination, on account of
a really stunning argument.

\subsection{Witness the sleight of hand!}
GHSZ continue.  ``The trouble becomes manifest, and an actual contradiction emerges, when we use $(11b)$ -- which until now has not been brought into play -- to obtain \vspace{.2cm}\\
\hspace*{4cm}$A_{\lambda}(\theta + \pi)B_{\lambda}(0)C_{\lambda}(\theta)D_{\lambda}(0) \ = \ 1\, \ \ \ \ \ \ \ \ \ \vspace{.2cm} \ (17)$\\
which in combination with Eq. $(12b)$ yields \vspace{.2cm}\\
\hspace*{4cm}$A_{\lambda}(\theta + \pi) \ = \ -A_{\lambda}(\theta)\vspace{.2cm}\, . \ \ \ \ \ \ \ (18)$

\noindent 
``This result {\it confirms} the sign change that we anticipated on physical grounds in EPR's program, but it also {\it contradicts} the 
earlier result of Eq. $(16)$ (let $\phi = \pi/2, \theta = 0)$.  We have thus brought to the surface an inconsistency hidden in premises (i)-(iv).''  
Their parenthetical suggestion means to let $\theta = 0$ in $(18)$ and to let $\phi = \pi/2$ in $(16)$.\\

YOW! \ \  What could be a more stunning and insightful demolition of the EPR premises? \\

Answer:  some calm logical thinking, and  the truth!  Well, what could be wrong here?

\subsection{A little more denken}

To begin with a startling observation, 
neither of the equations numbered $(11a)$ and $(11b)$ stands on its own.
Both of them are conclusions of a conditioning clause, an ``if clause''; and these two clauses
are quite evidently contradictory.  Yet GHSZ use these two conclusions in concert.
Equation $(11a)$ results from quantum theory applied to a situation in which
the four magnet angles are designed to meet the condition ``If $\phi_1 + \phi_2 - \phi_3 - \phi_4 \ = \ 0$''.
Equation $(11b)$ provides
the conclusion to the condition ``If $\phi_1 + \phi_2 - \phi_3 - \phi_4 \ = \ \pi$''.
If either of these conditions holds for an experiment under consideration then the other
cannot.  The two conditions can not
be instantiated at the same time, and they may not
be honoured at the same time.  ...  and neither may the the distinct conclusions they motivate.
It is surely not permitted to combine their equation $(17)$ with equation $(12b)$ to yield $(18)$.
Professors Greenberger, Horne, Shimony and Zeilinger discover the contradiction in the EPS suppositions that they
do {\it because they have introduced the contradiction into their analysis themselves! }   Full stop.\\

Compounding the misunderstanding this analysis has supported, empirical results from an actual experiment in a companion context of three photon polarizations was presented by Pan et al (2000) claiming to prove the inadequacy of a locally realistic model to account for them.  This was challenged by Aschwenden et al (2006) who proposed just such a model that improved the experimental explanation provided by a purely quantum theoretic model.  In fact, problems raised by the empirical programme engaged by Pan et al run even deeper than this.  To avoid a distraction here I comment only briefly (explicitly enough only for readers acquainted with details of this highly regarded publication) in an Appendix which follows the references for this present paper.\\

Returning to our assessment of the GHSZ article itself, recognizing the joint supposition of the contradictory 
premisses to $(11a)$ and $(11b)$ alone
should have startled any serious reader.  It should
surely not allow the acclaim that has been afforded to the spurious result of GHSZ.
However, the confusions in their argument run even deeper still.
There is much more that can be learned by
a continued pursuit of the ``troublesome preliminary result'' $(16)$ from which their supposed
discovery of a contradiction then deterred them.
We shall learn now that this very result itself derives from another serious error.  Less evident to a new reader,
perhaps, it is nonetheless shocking, having arisen from insufficient thinking combined with
 the use of casual notation.  Let's look into it.

\subsection{Completing the notation, and continuing pursuit of trouble }

I had initially been taken in by the GHSZ argument, and I shared their
puzzlement.  However, whereas they were concerned with the identical signs of a spin observed from opposite directions,
I wondered how might the angle $\phi_1$ at station $A$ equal both $2\phi$ and equal $0$ so as to instantiate equation $(16)$?  The  quantities
$A(2\phi_1)$ and $A(0)$ explicitly denote spin observations at station $1$ in two different experiments in which the orientation of the detection magnet differs.  In either of them the outcome of the experiment is random, specified by quantum probabilities, but allowing each of them to equal $-1$ or $+1$.  There is no requirement of any
sort that these outcomes need be identical in the two different experimental runs.
What could their equation $A(\pi) = A(0)$ mean?
It turns out that there is a clear way out of either conundrum, as we shall see.    We shall reconsider their
development of $(16)$ and find how to clarify the situation.  Sad to say it, we shall need to start at the beginning.\\

The GHSZ argument was proposed some fifty years after the recognition of particle entanglement had arisen among
quantum theorists.  This had made all of us aware that consideration of any aspect of particle behaviour observed at
station $1$ with a measurement labeled $A$ will typically depend on
both the settings of the angles {\it at 
all four stations}, and depend on the behaviours of the particles at these other stations as well.
%the other
%three stations, and depend on the behaviour of the particles observed there as well.  
I mention this because at
the very start of their argument, GHSZ casually denote the particular instantiations of their expectation equations
$(10a,b)$ by writing  \vspace*{.2cm}\\
\hspace*{4cm}If \ \ \ $\phi_1 + \phi_2 - \phi_3 - \phi_4 \ = \ 0 $\vspace*{.2cm}\\
\hspace*{4cm}then \ \ \ $A_{\lambda}(\phi_1)B_{\lambda}(\phi_2)C_{\lambda}(\phi_3)D_{\lambda}(\phi_4) \ = \ -1\,, \ \ \ \ \ \vspace{.25cm}(11a)$\\
\hspace*{4cm}If \ \ \ $\phi_1 + \phi_2 - \phi_3 - \phi_4 \ = \ \pi $\vspace*{.25cm}\\
\hspace*{4cm}then \ \ \ $A_{\lambda}(\phi_1)B_{\lambda}(\phi_2)C_{\lambda}(\phi_3)D_{\lambda}(\phi_4) \ = \ +1\,. \ \ \ \ \ \vspace{.2cm}(11b)$\\
\noindent Equations $(10a,b)$ had quite rightly designated the expectations in their concluding clauses (the left-hand-sides
 of the equations in their ``then'' statements) as  functions of four directional variables,
as in $E^{{\bf \psi}}(\hat{{\bf n}}_1, \hat{{\bf n}}_2, \hat{{\bf n}}_3, \hat{{\bf n}}_4)$.  Although I had
remarked about the cryptic form of this notation, their verbal description of
it did have the feature of recognizing that this is an
expectation of a general function of four variables, the directional vectors of the four magnet settings.
However,  in their instantiation equations $(11a,b)$ GHSZ blithely represent this product function
$ABCD(\phi_1,\phi_2,\phi_3,\phi_4)$  as a separable function of four independent variables standing alone
in singular directional settings: \ $A(\phi_1)B(\phi_2)C(\phi_3)D(\phi_4)$.  For the moment I say only,
``Beware!''.  But this misconstrual of the situation becomes even more abusive, and we shall need to
look into it deeply to
uncover the havoc  it has promoted.\\

A complete designation of the arguments of these spin measurement functions
$A, B, C$, and $D$ needs be made to denote fully
the experimental context in which the observations are made.  The
measurement which GHSZ denote as $A_{\lambda}(\phi_1)$ requires embellishment to $A_{\lambda}(\phi_1,\phi_2,\phi_3,\phi_4)$
if it is to denote an observation of the spin $A$ on the first
of four entangled particles.  The magnet angle at station 1 may
well have been set at $\phi_1$, but what were the directional angles for the spin detectors of the other three
particles with which the particle entering station $1$ is
entangled\.?  This surely makes a difference, as the distinction between
equations $(10a)$ and $(11a)$ makes clear.
On the one hand, one might might insist on the full experimental design notation $A(\phi_1,\phi_2,\phi_3,\phi_4)$ in
the identification of the spin measurement $A$ in any experimental run.  However this would amount to an ungainly
designation of a spin-product of four such measurements.  I am going to suggest and to follow henceforth a notation
that will use the subscript position under $A$ to designate the full angular context in which a measurement of $A$
at its station angle $\phi_1$ is made.  So we will write   $A_{(\phi_1,\phi_2,\phi_3,\phi_4)}(\phi_1)$ to designate the full angular context in which a measurement of $A$ at its
station angle $\phi_1$ might be made.  Remember that we are forgoing the unvarying GHSZ universal subscript of $\lambda$ on spin values $A,B,C,$ and $D$, except at times when I quote them exactly.\\

Using this notational convention we need to rewrite the GHSZ functional form equations $(11a,b)$ of the general
expectation result $(10a,b)$,  in the form  \vspace*{.2cm}\\
\hspace*{4cm}If \ \ \ $\phi_1 + \phi_2 - \phi_3 - \phi_4 \ = \ 0 $\vspace*{.2cm}\\
\hspace*{.3cm}then \ \ \  $A_{(\phi_1,\phi_2,\phi_3,\phi_4)}(\phi_1)B_{(\phi_1,\phi_2,\phi_3,\phi_4)}(\phi_2)C_{(\phi_1,\phi_2,\phi_3,\phi_4)}
(\phi_3)D_{(\phi_1,\phi_2,\phi_3,\phi_4)}(\phi_4) \ = \ -1\,, \ \ {\rm and} \ \ \  \ \ \vspace{.35cm}(11a)$\\
\hspace*{4cm}If \ \ \ $\phi_1 + \phi_2 - \phi_3 - \phi_4 \ = \ \pi $\vspace*{.25cm}\\
then \ \ \ $A_{(\phi_1,\phi_2,\phi_3,\phi_4)}(\phi_1)B_{(\phi_1,\phi_2,\phi_3,\phi_4)}(\phi_2)C_{(\phi_1,\phi_2,\phi_3,\phi_4)}
(\phi_3)D_{(\phi_1,\phi_2,\phi_3,\phi_4)}(\phi_4) \ = \ +1\,. \ \ \ \ \ \ \ \ \ \ \vspace{.2cm}(11b)$\\
\noindent Apologies for this complex notation, but we shall require it to air an egregious error in the GHSZ argument,  to which we now turn.\\

Appearing now {\it somewhat} less ungainly, the specific instantiation equations they enumerate as equations $(12a,b,c,d)$
would be designated by \vspace{.2cm}\\
\hspace*{2.4cm}$A_{(0,0,0,0)}(0)B_{(0,0,0,0)}(0)C_{(0,0,0,0)}(0)D_{(0,0,0,0)}(0) \ \ = \ \ -1 \vspace{.2cm} \ \ \ \ \ \ \ \ \ \ \ \ (12^*a)$\\
\hspace*{2.1cm}$A_{(\phi,0,\phi,0)}(\phi)B_{(\phi,0,\phi,0)}(0)C_{(\phi,0,\phi,0)}(\phi)D_{(\phi,0,\phi,0)}(0) \ \ = \ \ -1 \ \ \ \ \vspace{.2cm} \ \ \ \ \ \ \ \ (12^*b)$\\
\hspace*{2.1cm}$A_{(\phi,0,0,\phi)}(\phi)B_{(\phi,0,0,\phi)}(0)C_{(\phi,0,0,\phi)}(0)D_{(\phi,0,0,\phi)}(\phi) \ \ = \ \ -1 \vspace{.2cm} \ \ \ \ \ \ \ \ \ \ \ \ (12^*c)$\\
\hspace*{2cm}$A_{(2\phi,0,\phi,\phi)}(2\phi)B_{(2\phi,0,\phi,\phi)}(0)C_{(2\phi,0,\phi,\phi)}(\phi)D_{(2\phi,0,\phi,\phi)}(\phi) \ \ = \ \ -1 \vspace{.2cm}, \ \ \ \ (12^*d)$\\
for the station angle vectors that subscript the multiplicands in all of these products meet the condition that
$\phi_1 + \phi_2 - \phi_3 - \phi_4 = 0$, the condition under which $(11a)$ holds.\\
%for the components of the subscripted angle configuration vectors underlying each of these product designations meet the condition that
%$\phi_1 + \phi_2 - \phi_3 - \phi_4 = 0$
%, the condition under which $(11a)$ holds.\\

As an appeasement to your understandable hopes for a simpler notation,
I mention  here only that there will be some contexts in which
it will be sufficient to subscript a spin observation only by the value of $\kappa = \phi_1+\phi_2-\phi_3-\phi_4$, so to be
writing something like $A_{\kappa}(\phi_1) = A_0(0)$ in place of $A_{(0,0,0,0)}(0)$.  However in the four lines of $(12^*)$,
the experimental contexts are such that every spin observation would then be subscripted with $0$.  This would mean, for examples,
that $D_{(0,0,0,0)}(0)$ in $(12^*a)$ and $D_{(\phi,0,\phi,0)}(0)$ in $(12^*b)$, would both be designated by
 $D_0(0)$, whereas in fact they quite rightly designate very different things, these being the observation values of $D$ in two
 quite different experiments and experimental settings.  There is no assurance at all that
 they will instantiate at the same numerical
 value, as we shall now see.  GHSZ casually ignore this whole contextual matter about which their knowledge of quantum
 entanglement should have alerted them. They designate them both by $D_{\lambda}(0)$, and presume that they are always
 equal in any run of their experiment, feeling free to cancel them against one another when they do their algebra.
 Then they are briefly
 surprised by their equation $(16)$ which results.  However, as they carry on with the self-contradictory analysis we have discussed,
they then ignore the entire issue, surprising preliminary result and all.  To the contrary, we shall now trudge into these
 matters in great gory detail, but we shall first provide a clarifying assessment of the entire setup.

\subsection{Clarification via examination of the realm matrices}

For clarification of a central aspect of the situation relevant to all that follows,
my equation $(i)$ displays the ensemble of observation possibilities (which I call a ``realm matrix'') for an
observable spin vector arising in the conduct of a specific Stern-Gerlach
experiment on four entangled particles.  The observations are made on four particles in
a quantum state $|{\bf \Psi}\rangle$ corresponding to {\it any} specific
experimental design for which $\phi_1 + \phi_2 - \phi_3 - \phi_4 \ = \ 0$.
To repeat, this condition is a supposition (``If clause'') of GHSZ equation $(11a)$ which does not stand on its own without this clause.
Notice that the companion equation $(11b)$ relies on an alternative condition that $\phi_1 + \phi_2 - \phi_3 - \phi_4 \ = \ \pi$.
Contradictory to each other, both of these conditions cannot be satisfied in any specific experimental run, nor even in the
imagined runs of a gedankenexperiment on a specific single quartet of particles. The angle combination may equal either $0$ or $\pi$
in any exerimental setup,
but it cannot equal both at the same time. Nor can their two concluding
equations hold at the same time if we are to insist on discussion that honours the prescriptions of deductive logic.
It is a sad commentary on our times that we need to make explicit this proviso.  \begin{center}
	\noindent \hspace*{1cm}${\bf R}\left(\begin{array}{c}
	A_0(0) \\
	B_0(0) \\
	C_0(0)\\
	D_0(0)
	\end{array}\right) \ \ = \ \
	\left(\ \begin{array}{cccccccc}
	1 & 1 & 1 & \llap{$-$}1 & \llap{$-$}1 & \llap{$-$}1 & \llap{$-$}1 & 1 \\
	1 & 1 & \llap{$-$}1 & 1& \llap{$-$}1 & \llap{$-$}1 & 1 & \llap{$-$}1 \\
	1 & \llap{$-$}1 & 1 & 1 & \llap{$-$}1 & 1 & \llap{$-$}1 & \llap{$-$}1\\
	\llap{$-$}1 & 1 & 1 & 1 & 1 & \llap{$-$}1 & \llap{$-$}1 & \llap{$-$}1
	\end{array}\right)  \ \ \equiv \ \ {\bf R}_{-1}$  \ \ \ . \ \ \ \ \ \hspace*{1cm}$(i)$
\end{center}
In displaying this realm matrix, I denote the specific experimental quantity observation vector in its shorthand
form using the subscript $\kappa$ just mentioned: \ $[A_0(0), B_0(0), C_0(0), D_0(0)]^T$.
The subscripts $0 = \kappa$ on each of the spin observation values denotes the
contextual value of the angle combination pertaining to the experiment, and the arguments of the vector components
designate that each of the four angles $\phi_i$ is equal to $0$ radians.
 However, it would constitute no loss of generality to recognize this realm matrix of possibilities
 as pertinent to a spin observation vector
 resulting from any design that meets this four-angle condition, $\kappa = 0$.  \\

Since this restriction that  $\phi_1 + \phi_2 - \phi_3 - \phi_4 \ = \ 0$ implies via equation (9) that the expected spin-product $A_\lambda(0)B_\lambda(0)C_\lambda(0)D_\lambda(0)$ equals $-1$, it would be impossible to achieve
any four experimental spin results that allow the vector of
these multiplicands to imply a positive product.  The quantum probability weight on such observation vectors
must equal $0$, and such observation vectors would be impossible.
(If such an observation {\it were} seemingly made, the experimental quantum theorist would typically reject it as valid, and check the settings of the
direction angles of the four Stern-Gerlach magnets in the experimental run that generated it.)  
In the entangled state of the four-particle system specified by $|{\bf \Psi}\rangle$ in equation $(7)$,
the columns of this matrix exhaust the vector values of measurements that can arise from such an experiment.
The product of any of these eight ensemble column components equals $-1\,$.\\

Notice the concluding {\it definition symbol}, $(\equiv)$, introducing the denotation ${\bf R}_{-1}\;$
at the right end of equation $(i)$.  This is to distinguish it from a companion matrix to be
denoted by ${\bf R}_{+1}$ which specifies the
realm matrix corresponding to the possible outcomes of a {\it different experiment} in which the magnet angles satisfy
instead the alternative condition providing for $(11b)$, that $\phi_1 + \phi_2 - \phi_3 - \phi_4 \ = \ \pi\,$.
Again, without loss of generality,
an exemplar experiment would
generate an observable result $(A_\pi(\pi), B_\pi(0), C_\pi(0), D_\pi(0))^T\,$, with a realm matrix of columns
whose products all equal $+1\,$:
\begin{center}
\noindent \hspace*{1cm}${\bf R}\left(\begin{array}{c}
A_\pi(\pi) \\
B_\pi(0) \\
C_\pi(0)\\
D_\pi(0)
       \end{array}\right) \ \ = \ \
\left(\begin{array}{cccccccc}
1 & \llap{$-$}1 & \llap{$-$}1 & \llap{$-$}1 & 1 & 1 & 1 & \llap{$-$}1  \\
1 & \llap{$-$}1 & 1 & 1 & \llap{$-$}1 & \llap{$-$}1 & 1 & \llap{$-$}1 \\
1 & 1 & \llap{$-$}1 & 1 & \llap{$-$}1 & 1 & \llap{$-$}1 & \llap{$-$}1\\
1 & 1 & 1 & \llap{$-$}1 & 1 & \llap{$-$}1 & \llap{$-$}1 & \llap{$-$}1
             \end{array}\right)  \ \ \equiv \ \ {\bf R}_{+1}$  \ \ \ . \ \ \ \ \ \hspace*{1cm}$(ii)$
             \end{center}

\vspace*{.3cm}
The restrictions on the measurement vector possibilities embedded in the realm matrices
${\bf R}_{-1}$ and  ${\bf R}_{+1}$ derive from equation (9) of GHSZ, which specifies that
$E^{{\bf \psi}}(\hat{{\bf n}}_1, \hat{{\bf n}}_2, \hat{{\bf n}}_3, \hat{{\bf n}}_4) \ = \ -cos(\phi_1 + \phi_2 - \phi_3 - \phi_4)$.
At the two extreme angle restrictions we have entertained, that $\phi_1 + \phi_2 - \phi_3 - \phi_4$ equal $0$ or $\pi$,
this negative cosine value equals $-1$ and $+1$ respectively.  This is what restricts the measurement realms to be ${\bf R}_{-1}$ and
${\bf R}_{+1}$ in these extreme cases.  If the combination of experiment angles in this equation were to equal some other value, say, $\kappa \in (0,\;\pi)$,
then the realm  matrix of
possibilities for the measurements of the four electron spins would be the concatenation of these
two realms, $[{\bf R}_{-1}\ {\bf R}_{+1}]$.  With the recognition of this situation clearly in mind, we are ready to face the wall.

\subsection{Hitting a wall at equation $(16)$? Can we get there? Where {\it do} we arrive?}   % HERE HERE HERE

GHSZ regard their conclusion (16), that $A_\lambda(2\phi) = A_\lambda(0) =  const$ \ for any angle $\phi$, as surprising, for reasons they have well explained.
%\begin{center}
%$A_\lambda(2\phi) = A_\lambda(0) = $ const for all $\phi$.
%\end{center}
\noindent  Thinking that nonetheless this equation is not mathematically contradictory in itself, they
were sidetracked into the mistaken analysis that presumes jointly the contradictory suppositions of their
 full two-lined expressions of $(11a)$ and $(11b)$.  ``Finding'' a contradiction in their results, they quit.\\

Having recognized their error, we shall continue our investigation of their conundrum.  We have already
recognized that their equation $(16)$ derives from analysis that uses incomplete notation for describing the
situation.  Using our completed notation, we have embellished their statement of equations $(11)$ and $(12)$ to
the formulation in $(11^*a,b)$ and $(12^*a,b,c,d)$.  Let's now try to follow their line of argument using our complete notation, and see whether we will still hit the wall at their $(16)$.

\subsubsection{Heading toward the wall}

As a consequence of equalities $(12a)$ and $(12b)$ GHSZ obtained \vspace{.2cm}\\
\hspace*{2.8cm}$A(0)B(0)C(0)D(0) \ \ = \ \ A(\phi)B(0)C(\phi)D(0)\vspace{.2cm} $\ , \\
which allowed them to conclude\vspace{.2cm}\\
\hspace*{4.2cm}$A(\phi)C(\phi) \ = \ A(0)C(0) \vspace{.2cm}\ \ \ \ \ \ \ \ \ (13a)$, \\
through cancellation of $B(0)D(0)$ which appears identically on both sides of this
equation.  However, we find that writing the equations $(12a,b)$
in full notation that includes the contextual subscripts,
the equalities $(12^*a)$ and $(12^*b)$
obtain for us the equation \vspace{.25cm}\\
\hspace*{.5cm} $A_{(0,0,0,0)}(0)B_{(0,0,0,0)}(0)C_{(0,0,0,0)}(0)D_{(0,0,0,0)}(0) \ \ = \ \ \vspace{.3cm}\\
\hspace*{5cm} = \ \ A_{(\phi,0,\phi,0)}(\phi)B_{(\phi,0,\phi,0)}(0)C_{(\phi,0,\phi,0)}(\phi)D_{(\phi,0,\phi,0)}(0)\vspace{.23cm} \ .\ \ \ \ \ \ \ (13^*a)$\\
\noindent In full notation, the multiplicands that appeared identically as $B(0)D(0)$ on the left-hand sides of $(12a)$ and
$(12b)$, now appear as two quite different things in $(12^*a)$ and
$(12^*b)$, ... as they should!  For they represent the products of the $B$ and $D$ spins observed in two completely different
experiments:  $B_{(0,0,0,0)}(0)D_{(0,0,0,0)}(0)$ and $B_{(\phi,0,\phi,0)}(0)D_{(\phi,0,\phi,0)}(0)$.  \\

It is true
that the products of all four subscripted experimental settings on the two sides of this equation must equal $-1$.
However, there is no restriction at all on the product of any {\it two} of them.  Examining the realm matrix ${\bf R}_{-1}$ it is clear that the product  $B_0(0)D_0(0)$ may equal either $-1$ or $+1$
in any experimental run for which the directional angle setting specifies $\kappa = -1$.
Evaluate the componentwise products of rows $2$ and $4$ of equation $(i)$ to see that the elements of the
resulting product row alternate between $-1$ and $+1$.
In any two such runs there
is no assurance that the value will be the same.  We are not permitted to cancel the terms
$B_{(0,0,0,0)}(0)D_{(0,0,0,0)}(0)$ and $B_{(\phi,0,\phi,0)}(0)D_{(\phi,0,\phi,0)}(0)$
from the two sides of our equation $(13^*a)$.  They are relevant to the observations of two distinct experiments; and
either, both, or neither of them might be equal to $-1$. The spin-product combination $\kappa$ of all four
subscripted angles  specifies only
whether the expectation of the product of {\it all four} spin values must equal $-1$ or not.\\

The argument of GHSZ that obtains their equation $(13a)$ is specious.  It does involve a more sophisticated error than
their simple error of contradictory logic which has already been castigated.  But it is a seriously consequential
error nonetheless, an error of insufficient thought applied to casual notation.  
%I might mention here in passing that I suspect this type of error may run deeply through further supposed mysterious results of quantum theory.  I have found it arising in a very different context in another claim to a differentquantum mystery, which shall not deter us now.\\
%in an article by WASHINGTON U GUY (19??). He was one of two notable
%quantum theorists who had alerted me that my assessment of Aspect/Bell had become irrelevant due to the pathbreaking
%results of GHSZ.  That's when I began reading them and thinking.  Without further remark
%here, for now I will leave you to find that error for yourselves.\\
Continuing with this investigation, we find GHSZ repeat this error in their ``derivation'' of (13b), leading to a real
fiasco in their development of equations $(14a,b)$, their ``surprising result'' of $(15)$, and their ``preliminary
result'' of
$(16)$.  Their ``troublesome conclusion'' to it all, that $A(\pi)$ must equal $A(0)$, can now be dismissed as gibberish.
We never reach the wall at all!

\subsubsection{Where have we arrived?}

Well, what is to be made of the considerations of GHSZ?  We had followed them
through their equations $(9)$ and $(10)$, albeit our allusion to their use of cryptic notation.  We had proposed that
the conclusion of their analysis through $(10)$ would be expressed more clearly as
$E^{{\bf \psi}}[ABCD(\phi_1,\phi_2,\phi_3,\phi_4)] \ = \ -  cos(\phi_1 + \phi_2 - \phi_3 - \phi_4)$.  Using this complete  notation, the entanglement
among the four multiplicand spins is evident.  Moreover, it is
readily apparent that this conclusion displays a symmetry of the quantum spin-product expectation with respect
to rotations of the entire 4-ply Stern-Gerlach mechanism in the $(x,y)$ dimensions at the four observation stations,
and more! For as long as $t_1+t_2 = t_3+t_4$, it is evident that \vspace{.15cm}\\
\hspace*{1.5cm}$E^{{\bf \psi}}[ABCD(\phi_1+t_1,\phi_2+t_2,\phi_3+t_3,\phi_4+t_4)] \ = \ E^{{\bf \psi}}[ABCD(\phi_1,\phi_2,\phi_3,\phi_4)], \vspace{.15cm}\hspace{1cm}(iii)$\\
because the value of $\kappa$ associated with both of these 4-angle designs is identical. Let's think.

%*****\\
%  THESE NEXT LINES WEERE JUST WRITTEN TO TEASE GHSZ ABUT NOT READING CONDITIONAL CLAUSES
%\noindent If $r_1+r_2 = r_3+r_4$, then $E^{{\bf \psi}}[ABCD(\phi_1+r_1,\phi_2+r_2,\phi_3+r_3,\phi_4+r_4)] \ = \ E^{{\bf \psi}}[ABCD(\phi_1,\phi_2,\phi_3,\phi_4)]$.\ \ \vspace{.15cm}(iii) \\

%\noindent $E^{{\bf \psi}}[ABCD(\phi_1+r_1,\phi_2+r_2,\phi_3+r_3,\phi_4+(r_1+r_2-r_3)] \ = \ E^{{\bf \psi}}[ABCD(\phi_1,\phi_2,\phi_3,\phi_4)]\ \ \ \forall {\bf r}_3.$\ \ \vspace{.15cm}(iii) \\
%*****\\
%\indent For as long as $t_1+t_2 = t_3+t_4$, it is evident that \vspace{.15cm}\\
%\hspace*{1.5cm}$E^{{\bf \psi}}[ABCD(\phi_1+t_1,\phi_2+t_2,\phi_3+t_3,\phi_4+t_4)] \ = \ E^{{\bf \psi}}[ABCD(\phi_1,\phi_2,\phi_3,\phi_4)], \vspace{.15cm}\hspace{1cm}(iii)$\\
%because the value of $\kappa$ associated with both of these 4-angle designs is identical.
%\noindent I cringe when numbering this as equation $(iii)$ because the statement is too long to print on a
%single line, and I fear that GHSZ will consider the concluding equation to stand on its own.  So for them
%I will rewrite it here more densely, but in a form that does not require an ``if'' clause, and number it for them
%as $iii')$.  Smile! \vspace*{.2cm}\\
%\noindent $E^{{\bf \psi}}[ABCD(\phi_1+r_1,\phi_2+r_2,\phi_3+r_3,\phi_4+(r_1+r_2-r_3))] \ = \ E^{{\bf \psi}}[ABCD(\phi_1,\phi_2,\phi_3,\phi_4)]\ \ \ \forall {\bf r}_3.\ \ (iii')$\\ %\vspace{.15cm}
% OK, back to statement $(iii)$.  Sorry for the abuse.  I am teasing you.  :)\\

\subsubsection{Recognizing rotational as well as permutation symmetry}

Algebraically, two types of symmetry are evident in the QM-motivated stipulations of equations $(8)$ and $(9)$.  These show that
the expected 4-spin-product is a function only of the angle combination
$\phi_1 + \phi_2 - \phi_3 - \phi_4\,$.
Rotational symmetry of the experimental conditions would be exhibited in the transformation of the vector of
angles $({\phi}_1,{\phi}_2,{\phi}_3,{\phi}_4)$ by the addition of any vector of
constants $(t,t,t,t)$,
which would surely preserve the angle combination $\kappa$.  In fact, preservation would continue under the addition of any angle vector ${\bf t}_4$ for
which $t_1+t_2 = t_3+t_4$.  This general condition allows permutations of either or both of
$\phi_1$ with $\phi_2$ and/or
$\phi_3$ with $\phi_4$ in the specification of ${\bf \kappa}$.  It would also allow the permutation of any pair $(\phi_1,\phi_2)$ with $(\phi_3,\phi_4)$,
because the cosine of any angle is identical to the cosine of the negative
 of that angle. Geometrically, rotational symmetry would allow rotation of the $(x,y)$ planes containing the directional vectors
$(\hat{{\bf n}}_1, \hat{{\bf n}}_2, \hat{{\bf n}}_3, \hat{{\bf n}}_4)$ in Figure $1$
around the $(x,y)$ axes orthogonal to the $-z \leftrightarrow z$ axes, all to the same degree. Permutation symmetry would allow the exchange
of the $(x,y)$ axes of directional designations between stations $1$ and $2$ or between $3$ and $4$, .... or an exchange
of these axis systems of stations $(1,2)$ with those of $(3,4)$.   Moreover, the invariance with respect to
$t_1+t_2 = t_3+t_4$ is even richer than all these symmetries, because it allows the twisting of the $(x,y)$ axes at
station $(1)$ and station $(2)$ in any ways one wishes, just so long as one concomitantly would twist the axes at station $(3)$ and station $(4)$
correspondingly, ensuring that the {\it sum} of the two latter twisting degrees equals
the {\it sum} of the twisting degrees of the the former two. \\
%  Bring on Chubby Checker!  Twisting the night away!\\

%Enough mirth!  This is quantum physics!  Heavy stuff.\\

A final comment on this situation is in order before we consider a Bell inequality formulation in the context of the 4-dimensional GHSZ
experiment proposal.  Notice that the QM-motivated
invariance equation $(iii)$ specifies the invariance of {\it the expectation of the product} of the four spin
observations with respect to all these forms of twisting any initial reference 4-angle configuration.
However, equation $(iii)$ does {\it not} specify  invariance of  the
component multiplicands themselves that generate the product.
Each time we suggest a twisting of the four Stern-Gerlach
magnet orientations in the $(x,y)$ planes at the four stations, however we propose to do it,
we are considering a new distinct run of the 4-ply spin
experiment of GHSZ.  Even in the special cases that the initial reference
configuration specifies $\kappa$ equal to $0$ or $\pi$, the invariance pertains to
the product of the multiplicand spin observations, not to the multiplicand spin values themselves.\\

This is what quantum theory tells us about the 4-dimensional experiments that GHSZ propose.  We have not yet
concerned ourselves with the claims regarding local realism and hidden variables, the issues underlying the
type of experiment they devised.\\

Now why did they propose their analysis at all?  They were attempting to study a higher dimensional example of the
Bell gedankenexperiment, to see how the inequality might fare.  Before they were deterred and
sidetracked from this concern by their errors of logic and of casual notation, they were trying to formalize the
implications of the principle of local realism for a four-dimensional experiment.  As it turns out, the formalization
can be constructed without confronting any contradiction at all, but we must be a little bit careful.
A preliminary airing of a few issues will conclude my analysis here.

\section{Expanding the CHSH/Bell formulation to four dimensions}

Remember that Einstein's principle of local realism pertains to situations that lie outside the scope of quantum
theory.  In the context of a GHSZ 4-ply experiment on two pairs of electrons, his assertion concerns the outcome of
other experiments
that one might consider conducting in tandem with the 4-ply experiment
that one does conduct.  However, these others are precluded from concomitant execution by the one that is conducted.
According to the general form of the uncertainty principle, quantum theory expressly renounces
 any scope for making proclamations about the joint result of such incompatible ventures.
 It can and does make probabilistic statements separately about the results of any one of the 4-ply experiments that may be considered.
No worries.  We can at least {\it think} about the prospect of such an impossible joint venture, and assess the relevance to its conduct of
what QM does say. Many many people have.  Well, what
would we need to think about if we were to think about such vagaries in the context of the GHSZ experiment?
At least we now have a vocabulary and a syntax of language to talk about what we are thinking.\\

Suppose, for example, that the outcome of a specific run of the experiment characterised by a 4-angles setting of
$({\phi}_1,{\phi}_2,{\phi}_3,{\phi}_4)$ for which $\kappa =
{\phi}_1+{\phi}_2-{\phi}_3-{\phi}_4 = 0$, yields the observed spin results \vspace{.2cm}\\
\hspace*{.3cm} $[A_{(\phi_1,\phi_2,\phi_3,\phi_4)}(\phi_1), B_{(\phi_1,\phi_2,\phi_3,\phi_4)}(\phi_2), C_{(\phi_1,\phi_2,\phi_3,\phi_4)}(\phi_3), D_{(\phi_1,\phi_2,\phi_3,\phi_4)}(\phi_4)] \ = \vspace{.2cm}\ (+1,-1,-1,-1)$\ .\\
Notice just to begin that the spin-product of the components of this supposed result vector is $-1$, as is required for a setup specifying this value
of $\kappa = 0$.
%However, the value of $\kappa$ for this design might equal any other number in the interval $(0,1]$.
Now about this observed result vector,
the principle of local realism would say, among other things, that if we were to conduct concomitantly  a companion experiment {\it on this
same quartet of electrons} at slightly adjusted angle settings of the form $({\phi}_1,{\phi}_2+t,{\phi}_3,{\phi}_4+t)$, for an example that preserves
the value of $\kappa = 0$,  then the result vector for this
alternative experiment would only have to satisfy the  \vspace{.25cm} form
\hspace*{.1cm} $[A_{({\phi}_1,{\phi}_2+t,{\phi}_3,{\phi}_4+t)}(\phi_1), B_{({\phi}_1,{\phi}_2+t,{\phi}_3,{\phi}_4+t)}(\phi_2+t), C_{({\phi}_1,{\phi}_2+t,{\phi}_3,{\phi}_4+t)}
(\phi_3), \vspace{.25cm} D_{({\phi}_1,{\phi}_2+t,{\phi}_3,{\phi}_4+t)}(\phi_4+t)] \\
\hspace*{12.5cm} \ = \vspace{.25cm}\ (+1,x,-1,x)$,\\
where the value of $x$ might equal either $-1$ or $+1$.  Either value would ensure that the product of the component observations equals $-1$ as required.
The claim of local realism is that the behaviour of these electrons at stations $1$ and $3$ would remain the same in any other run in any other setting
of the angles at stations $2$ and $4$, and moreover that the prescriptions of quantum probabilities must be preserved.
If the twists
of the magnets at $2$ and $4$ would destroy the value of $\kappa$ in the concomitant run,
then the spin values at $1$ and $3$ {\it for this same pair of electrons} would still remain the same according to local realism,
but the concomitant spin values at $2$ and $4$ might enjoy more liberty.  This would depend on the value of $\kappa$ associated with
the adjusted magnet angles $\phi_2'$ and $\phi_4'$. If the value of $\kappa$ for the adjusted magnet angles were not equal to
either $0$ or $\pi$ then the spin values at stations $2$ and $4$ would be permitted to alternate in sign.  This would be allowed
because the joint distribution of the component observations $(A,B,C,D)$ would be
different in the new angle settings according to quantum theory, and even the QM expectation
of the of the product $ABCD$ would then be different.  It would be only the spin values at angles unchanged that
would be forced to remain the same in the two concomitant runs on the same pairs of photons.\\

  It is because the electrons are entangled that the specification of the consequences of local realism
must take all four angle settings into account in the higher dimensional context.
Both of these distinct spin vector observations $(1,1,-1,1)$ and $(1,-1,-1,-1)$ would ensure that the product of the spin values would equal $-1$
when the angles $\phi_2$ and $\phi_4$ are both twisted by the same incremental value $t$.
So there are two possible observation vectors for this concomitant experiment that would satisfy this proclamation of local realism, but there are many
more impossibilities that would not.  For examples, an observation vector $(1,1,-1,-1)$ would be impossible in the twisted case, since it does not even respect the tenets of
quantum theory. (Remember we are presuming that $\kappa = \phi_1+\phi_2+\phi_3-\phi_4 = 0$, so the product of the four observations
must equal $-1$.) Further, an observation of $(1,-1,1,1)$ would be impossible as well,
not being in accord with local realism.  Since the value of $C_{(\phi_1,\phi_2,\phi_3,\phi_4)}(\phi_3)$ is proposed as $-1$ in the contextual angle setting initially proposed, local
realism would require it to equal $-1$ in the second setting as well for which $\phi_3$ is the same.
This second disallowed result would not accord with the principle of local realism, even though the proposed component product equals $+1$.
  Let's think some more.\\

%Wow!  Well, what a thing to say! Can you ever imagine saying anything like that?  Let's think about what this is that it says.\\

The principle of local realism specifies that while the results of a quantum experiment may well be stochastic, assigned
various probabilities according to the tenets of quantum theory, if the results at the observation stations $1$ and $3$ in any actual
experimental instance equal $+1$ and $-1$, say, then these results would have had to be the same in their instantiation in
other circumstances of their companion angles as well.  If these two electrons
had been arriving at their stations in tandem with the other
two electrons arriving at their observation stations at different angles than they had, then the results at these stations $1$ and $3$ would need to remain the same.\\

I shall now outline a construction of a Bell quantity in CHSH form that would be appropriate to the four dimensional problem of GHSZ.
I believe I have delineated the relevant details allowing how it might be done. As befits the pleasures of quantum physical theorizing,
there will be intrigue involved.  What I plan to do is to design  the performance a GHSZ experiment on a single quartet of doubly paired electrons
under $16$ distinct conditions.  I shall use alternate
magnet orientations at the opposing stations $1$ and $3$ and at the opposing stations $2$ and $4$ that mimic the four polarization settings
used by Aspect in his experiments with single pairs of photons.  You should recognize that experiments at stations $1$ and $3$ on their
own (ignoring the parallel experiments at stations $2$ and $4$) would replicate the simpler experiments of Aspect/Bell at stations $A$ and $B$.

\subsection{Setting the context }

We shall now embark on details that require us to clarify two peculiarities of the GHSZ depiction of their 4-ply experiment, which we have reprinted as
our Figure $1$.  To begin, examine the two  $(x,y)$ axis portrayals that are situated perpendicular to the ${\bf z}^+$ and ${\bf z}^-$ axes
at which a pair of electrons would enter stations $1$ and $3$.  On their own (ignoring for now the parallel axis systems relevant to stations $2$ and $4$)
they
constitute a replication of the planes in which the angling
pair of polarizers were situated in the Aspect/Bell problem.  In the GHSZ situation, the electron passing the S-G magnet at station $1$ enters the complex from the source
of its propagation, and the electron passing the magnet in the $(x,y)$ plane at station $3$ enters the complex from the same source.  Because this propagation source is
between the two station planes, the electrons are traveling in the opposite directions, these being ${\bf z}^+$ and ${\bf z}^-$.
As the two $(x,y)$ planes they enter are
designed to be identical, the plane depicted at station $3$ in the GHSZ Figure is displayed incorrectly,  as a reflection (about the vertical x-axis) of
its true orientation.    Visualize the directional vector depicted at station $3$ but with the y-axis
rotated by $\pi$ radians about the x-axis.  Think about it as you view the axis from behind, entering from the propagation source.
The depicted vector sweeps an angle of $\pi+\phi/3$ from the x-axis rather than $\phi_3$ as labeled.  You will get it.\\

Fair enough, it is only a picture, and we can easily adjust our understanding of the situation for this miscue.  For the algebraic derivation of the expectation formula
$E^{{\bf \psi}}[ABCD(\phi_1,\phi_2,\phi_3,\phi_4)]$ in their Appendix F seems to be appropriate to the real orientations of the $(x,y)$ planes at each station.
In the orientations displayed, the GHSZ Figure has the
benefit that it suggests correctly that the $(x,y)$ axis entered by each electron as it passes the magnet appears identical.  It would only be the direction of the S-G magnets
that might vary from station to station.\\

A second peculiarity of the displayed Figure merits mention.  Notice that at all stations the $(x,y)$ axis is depicted with the x-axis being vertical and the y-axis being horizontal.  This is unusual relative to
the common arbitrary orientation of these axes in which the x-axis runs horizontally from left to right as the value of $x$ increases,
while the y-axis runs vertically from bottom to top.
There is nothing improper about this if the situation is recognized for what it is.  The upshot of their non-standard convention relative to the sizes of the
angles $\phi_1,\phi_2,\phi_3,\phi_4$ can be understood by considering
the angle pairings specified by the directional vectors ${\bf a,b,a'}$, and ${\bf b'}$ in the original Aspect experiment.  Using the standard convention of
the directions of the $(x,y)$ plane, it is clear that the directional angles of Aspect's polarizers  would be $7\pi/16, 5\pi/16, 3\pi/16$, and $\pi/16$.  These would make the four angles
Aspect denoted by $({\bf a},{\bf b}), ({\bf a},{\bf b'})$, and  $({\bf a'},{\bf b}), ({\bf a'},{\bf b'})$ equal to $-\pi/8, -3\pi/8, \pi/8$, and $-\pi/8$ which I have used in my
analysis of the Aspect/Bell conundrum (Lad, 2020b).  (Remember that the angle denoted by $({\bf a},{\bf b})$ for example, is the angle passing between the directional vector ${\bf a}$
and ${\bf b}$.)
% Reading the situation according to the GHSZ depiction, the numerical values of these angles are the same, but one needs to twist your head around the back of their
%axis system and then cock it at a right angle to understand the situation.  So be it.
I think we can be clear in proceeding now with some numerical investigations.

\subsection{Now pursuing the gedankensetup}

I shall now provide an analysis of sixteen simultaneous gedankenexperiments at which the two pairs of electrons are sent to the four stations, concomitantly toward two different magnet angles
at each station.  I shall be brief, merely reporting the results of the analysis at each stage of its development rather than providing every interim detail.\\

The two chosen angles at each station $i$ will be designated as $\phi_i$ and $\phi_i'$.  According to the principle of local realism, we presume that any specific
instantiation of a spin observation at a station at a specific angle would arise as the same, no matter which other three angles were
engaged by the electrons at the other stations.  This is despite
the fact that the QM joint probability distribution for the outcome of all four observations (and for their product)
clearly depends on the settings of all
four magnet angles.  This is exactly the principle we followed in generating
results of the simpler gedankenexperiment on single photon pair in the framework of Aspect/Bell.  Identifying the
ensemble of possible observations (the realm matrix) for the eight spin results exhibited in the sixteen experiments involving eight arbitrary angles, we would describe it
in transposed form as
${\bf R}[A(\phi_1),B(\phi_2),C(\phi_3),D(\phi_4),A(\phi_1'),B(\phi_2'),C(\phi_3'),D(\phi_4')]^T$ and would recognize the matrix columns as composing
the cartesian product of eight component vectors, $\{-1,+1\}^8$.\\

Now beneath each column of this $8 \times 256$ matrix, we would exhibit the vector of 4-product spin results that corresponds to this possibility at each of the sixteen scenarios
of the concomitant gedankenexperiments.  Sequentially, this vector would designate the ordered 4-spin-products  using none, one, two, three, and then four of the ``alternate'' primed magnet
angles $\phi_i'$ rather than the base
orientation at angle $\phi$.  Specifically, this section of the full realm matrix would be designated as follows, with the values of $\kappa$ and of the QM-motivated expectation values
relevant to each component product shown to its right. \vspace{.25cm}\\
${\bf R}[A(\phi_1)B(\phi_2)C(\phi_3)D(\phi_4)\vspace{.03cm}\hspace{4.2cm} \ \ \ \pi/4\hspace{1.67cm} \ \ \hspace{1.5cm}-1/\sqrt{2}\\
\hspace*{.4cm}A(\phi_1')B(\phi_2)C(\phi_3)D(\phi_4)\vspace{.03cm}\hspace{4.2cm} \ \ \ \ \ 0\hspace{1.67cm} \ \ \ \hspace{1.5cm}\ \ \ \ -1\\
\hspace*{.4cm}A(\phi_1)B(\phi_2')C(\phi_3)D(\phi_4)\vspace{.03cm}\hspace{4.2cm} \ \ \ \ \ 0\hspace{1.67cm} \ \ \ \hspace{1.5cm}\ \ \ \ -1\\
\hspace*{.4cm} A(\phi_1)B(\phi_2)C(\phi_3')D(\phi_4)\vspace{.03cm}\hspace{4.2cm} \ \ \ \pi/2\hspace{1.67cm} \ \ \hspace{1.5cm}\ \ \ \ \ \ \ 0\\
\hspace*{.4cm}A(\phi_1)B(\phi_2)C(\phi_3)D(\phi_4')\vspace{.03cm}\hspace{4.2cm} \ \ \ \pi/2\hspace{1.67cm} \ \ \hspace{1.5cm}\ \ \ \ \ \ \ 0\\
\hspace*{.4cm}A(\phi_1')B(\phi_2')C(\phi_3)D(\phi_4)\vspace{.03cm}\hspace{4.0cm}  \ -\pi/4\hspace{1.67cm} \ \ \hspace{1.5cm}-1/\sqrt{2}\\
\hspace*{.4cm}A(\phi_1')B(\phi_2)C(\phi_3')D(\phi_4)\vspace{.03cm}\hspace{4.2cm} \ \ \ \pi/4\hspace{1.67cm} \ \ \hspace{1.5cm}-1/\sqrt{2}\\
\hspace*{.4cm}A(\phi_1')B(\phi_2)C(\phi_3)D(\phi_4')\vspace{.03cm}\hspace{4.2cm} \ \ \ \pi/4\hspace{1.67cm} \ \ \hspace{1.5cm}-1/\sqrt{2}\\
\hspace*{.4cm}A(\phi_1)B(\phi_2')C(\phi_3')D(\phi_4)\vspace{.03cm} \ \ , \hspace{.9cm} {\rm for \ which} \hspace{.3cm}\ \ \kappa \ = \ ( \pi/4   ) \ \ \ \ \ {\rm and} \ \ \ \ \ E \ \ = \ \ ( -1/\sqrt{2}   )\\
\hspace*{.4cm} A(\phi_1)B(\phi_2')C(\phi_3)D(\phi_4')\vspace{.03cm}\hspace{4.2cm} \ \ \ \pi/4\hspace{1.67cm} \ \ \hspace{1.5cm}-1/\sqrt{2}\\
\hspace*{.4cm}A(\phi_1)B(\phi_2)C(\phi_3')D(\phi_4')\vspace{.03cm}\hspace{4.2cm} \ \ 3\pi/4\hspace{1.67cm} \ \ \hspace{1.5cm}+1/\sqrt{2}\\
\hspace*{.4cm} A(\phi_1')B(\phi_2')C(\phi_3')D(\phi_4)\vspace{.03cm}\hspace{4.2cm} \ \ \ \ \ 0\hspace{1.67cm} \ \ \ \hspace{1.5cm}\ \ \ -1\\
\hspace*{.4cm}A(\phi_1')B(\phi_2')C(\phi_3)D(\phi_4')\vspace{.03cm}\hspace{4.2cm} \ \ \ \ \ 0\hspace{1.67cm} \ \ \ \hspace{1.5cm}\ \ \ -1\\
\hspace*{.4cm} A(\phi_1')B(\phi_2)C(\phi_3')D(\phi_4')\vspace{.03cm}\hspace{4.2cm} \ \ \ \pi/2\hspace{1.67cm} \ \ \hspace{1.5cm}\ \ \ \ \ 0\\
\hspace*{.4cm}A(\phi_1)B(\phi_2')C(\phi_3')D(\phi_4')\vspace{.03cm}\hspace{4.2cm} \ \ \ \pi/2\hspace{1.67cm} \ \ \hspace{1.5cm}\ \ \ \ \ 0\\
\hspace*{.4cm}A(\phi_1')B(\phi_2')C(\phi_3')D(\phi_4')]\hspace{4.2cm}\ \pi/4\hspace{1.67cm} \ \ \hspace{1.5cm}-1/\sqrt{2}$\ \ . \vspace{.25cm}\\

GHSZ quite rightly derive the QM-motivated expectations for each one of these 4-spin-products via the specification \vspace{.25cm}

\hspace*{2cm}$E[A(\phi_1^*)B(\phi_2^*)C(\phi_3^*)D(\phi_4^*)]\  = \ -cos(\phi_1^*+\phi_2^*-\phi_3^*-\phi_4^*)$\ .\\

Having studied the simpler problem of Aspect/Bell, it is now no surprise to find that the realm matrix so generated for the sixteen 4-spin-products
does not display $256$ distinct
columns, but rather merely $32$.  What this tells us, is that if we can identify any five rows of this $16 \times 256$ matrix whose columns exhaust
the cartesian product $\{-1, +1\}$ then the remaining rows will have been identified to be functionally related to them.  It turns out that there are
many such combinations of five rows that will do this.  Of the $^{16}C_5 = 4368$ available for such choices of five rows, $2688$ of them provide a functional
relation.  In fact, all of these functional relations hold among the sixteen components of any column vector in the realm.  Although this is quite
a step up in the complexity of the functional restrictions among the 4-spin-products of a gedankenexperiment, it is structurally no different than
the four simultaneous functional relations that bound up the 2-polarization-products in the simpler context of the Aspect/Bell problem. \\

Let's stop for a moment to think about what this means: the conditional distribution for the other eleven 4-spin-products given any of these five
is degenerate on the function values they stipulate.   Ostensibly, we are searching for a QM-motivated joint probability distribution for the
results of all sixteen 4-spin-product quantities in the GHSZ gedankenexperimental design.  Of course the general uncertainty principle tells us
that there is no such joint distribution, because quantum theory says nothing about the joint result of any two such incommensurable 4-products.
However, if there were such a distribution then it could be factored as the product of the marginal joint distribution for any five of them times
the conditional distribution for the other eleven given these five.  Since this conditional distribution relative to any of the $2688$ function-producing
choice of five from sixteen is degenerate, the joint distribution for all sixteen would resolve to a specification of the joint distribution for
these five.  However quantum theory does not uniquely identify such a distribution.  What it does do however, is specify the expectations of the products
of any four spin values that might be observed in an experiment in any 4-angle setting.  Along with the the summation constraint, the five such product expectation values places six linear constraints on the probabilities for the
thirty-two constituents of the partition underlying the joint distribution.  It is precisely this situation to which the computational structure of
linear programming problems applies.\\

Well what should be the objective function of such a linear programming problem\.?  We have seen that the Bell quantity ``$s$'' in the Aspect/Bell analysis of the CHSH formulation 
 is a linear combination of the Expected products of all four components of the gedankenexperiment involved.  The coefficients
identifying the linear combination are all either $+1$ or $-1$.  So, consider as the objective function of the linear programming problem we might
formulate for the double electron pair experiment as a similar linear combination of the sixteen 4-spin-products we are considering.   Five linear constraints on the spin-products for the function domain settings plus the summation constraint would 
leave $26$ dimensions of freedom in the argument vector of the objective function.  The linear programming computation would find the vectors ${\bf q}_{32}^{min}$
and ${\bf q}_{32}^{max}$ that yield the extreme values of this linear combination. \\

To compute a complete solution to the gedanken problem we would need to compute $2688$ pairs of LP problems ({\it min} and {\it max}) to find all the
vertices of the polytope of probability distributions that quantum theory would allow as its solution.  Of course, as in the corresponding Aspect/Bell
solution, many of these would be duplicates.   At any rate, we would arrive at a larger sized solution of the quantum theory specified polytope
of probability distributions that is structurally similar to the solution we found in the case of Aspect/Bell.   End of story. \\

% One notional
%embellishment would be required to identify the context in which the spin at a particular station is observed in the 8-ply gedankenexperiment.  Rather
%than subscripting the measurement $A(\phi)$ merely by the four angles with which we observe it when we execute a real 4-ply experiment of the
%GHSZ design on two pairs of electrons, the outcome of the gedankenexperiment at each station would need be subscripted by all eight angles settings
%involved when all sixteen of the 4-spin-product results are observed for two pairs of electrons.   \\

%Now applying the QM specifications of Expectation values for each 4-product among the five so identifies

%blah blah blah \\

The only concluding remark is that every one of the feasible expectations for the objective function would respect the bounds specified by any appropriate Bell-type inequality.

\section{Technical conclusion}

Quite to the contrary of the GHSZ conclusion that the premises of EPR pose a contradiction to a quantum experiment involving four (and even only three)
particles, we can conclude that such experiments indeed do allow the premises of EPR, appropriately identified.  Furthermore, the conditions of such an
experiment can easily exemplify well
the EPR premise of ``Perfect correlation'', if desired, in both the case of conditions of $(11^*a)$ and that of $(11^*b)$. ... though not at the same time!  Smile.

\section{Concluding comments}

The results of this discussion have laid bare the claims of GHSZ who denigrate the logical consistency the EPR version of  ``local realism''.  However, I would not like a reader to think that I am in any way an advocate of the EPR construal of the situation either. From a larger perspective, I find it embedded in a view of physical experience that is seriously out of date.   My investigations of the past five years have been oriented quite narrowly, to a resolution of the conundrum posed by
the purported violation of Bell's inequality in the theory of quantum mechanics in any of the forms in which it has been promoted.  At a
deeper level, I conclude that the so-called mysterious properties of the quantum world, involving a structure of ``quantum probabilities''
which inhere different structures than the mundane probabilities of the world at the classical scale, have been misconstrued as well.  A discussion of
larger implications relevant to a reconstruction of physical theory awaits a confirmation of my narrow concerns. \\

As to the definitive empirical work of Hensen et al. (2016), I do not doubt their experimental results.  However, their statistical analysis suffers from the
same mistake as does that of Aspect.  Readers still impressed by it might review Section 6 my manuscript (Lad, 2020) on the mistaken violation of Bell's inequality.

\section*{Acknowledgments}

I am indebted to several unnamed physicists who have graciously
engaged with me in intense and detailed discussion over several years of any number of technical issues about which we have disagreed.  I have been challenged several times to address the GHSZ construct by honourable
colleagues who have been dismayed by my assessment of Aspect/Bell.  Thus, I now challenge you with this note.  My sincere thanks to Mike Ulrey who first alerted me to the intrinsic interest of Bell's inequality, to Duncan Foley who has supported my investigations as a scrupulous reader throughout, and to Diego Molteni who has provided both relevant background and collegial discussion.  This article has been published in an edited and revised format in an issue of {\it Entropy} 2020, 22, 0759; doi:10.3390/e22070759.  The published version uses a completely different equation numbering system, and a non-informative numbering system to reference articles. The version here also clarifies the text in a few instances.  Furthermore, in the published version the contents of the  Appendix that appears here following the References is inserted right into the text.

\section*{References}

\noindent {\bf Aschwanden, M., Philipp, W., Hess, K., and Barraza-Lopez, S.} (2006) {\it Quantum \\
	\indent Theory, Reconsideration of Foundations}, 3, AIP Conference Proceedings, 437-446.\\

\noindent {\bf Aspect, A.} (2002) Bell's Theorem: the naive view of an experimentalist,
{\it Quantum [Un]speakables \\
\indent -- from Bell to Quantum Information}, R.A. Bertlmann and A. Zeilinger (eds.),
Springer. \\

\noindent {\bf Brunner, N., Cavalcanti, D., Pironio, S. and Scarani, V.} (2014) Bell nonlocality,
{\it Reviews \\
	\indent of Modern Physics},  {\bf 86}(2), 420-478. \\

\noindent {\bf Einstein, A., Podolsky, B, and Rosen, N.} (1935) Can quantum mechanical
description of
\indent physical reality
be considered complete? {\it The Physical Review}, {\bf 47}, 777-780.  \\

\noindent {\bf Greenberger, D.M., Horne, M.A., Shimony, A., and Zeilinger, A.} (1990) Bell's \\
\indent Theorem without inequalities,
{\it American Journal of Physics},  {\bf 58}(12), 1131-1143. \\

\noindent {\bf Greenstein, G., and Zajonc, A.} (2006) {\it The Quantum Challenge: modern research on \\
	\indent foundations of QM}, Second edition, Sudbury, Mass: Jones and Bartlett\\

\noindent {\bf Hensen, B., Bernien, H, Dr\'{e}au, A.E., and a list} (2015) Loophole-free Bell inequality \\
\indent violations using electron spins
separated by 1.3 kilometers,
{\it Nature},  {\bf 526}, 682-686. \\

\noindent {\bf Jaeger, G.} (2009) {\it Entanglement, Information, and the Interpretation of Quantum Mechanics}\\
\indent 
Berlin, Heidelberg: Springer-Verlag.\\

\noindent {\bf Lad, F.} (2020a) Violation of  Bell's inequality:  a misunderstanding based on a mathematical \\
\indent error of neglect, 
Available at ResearchGate.\\

\noindent {\bf Lad, F.} (2020b) Resurrecting the principle of local realism, and the prospect of supplementary \\
\indent variables, 
Available at ResearchGate.\\

\noindent {\bf Mermin, N.D. and Schack, R.} (2018) Homer nodded: von Neumann's surprising oversight,\\
\indent {\it Foundations of Physics},  {\bf 48}(9), 1007-1020. \\

\noindent {\bf Pan, J-W., Bouwmeester, D., Daniell, M., and Zeilinger, A.} (2000) Experimental \\
\indent entanglement purification of arbitrary unknown states,  {\it Nature},  {\bf 423}, 417-422. \\

\noindent  {\bf von Neumann, J.} (1932, English translation 1955)
{\it Mathematical Foundations of Quantum\\ \indent Mechanics}, R. Beyer [tr], Princeton University Press. \\

%\noindent {\bf Look up the Washington U guy and his article}\\

\noindent {\bf Wiseman, H.} (2015) Death by experiment for local realism,
{\it Nature},  {\bf 526}, 649-650.

%\vspace{2mm}\\

\section*{Appendix:  A comment on the claims of Pan et al to empirical evidence}
The following terse comment is meant exclusively for any reader who is familiar with the setup, the notation, and the details of the respected article of Pan et al (2000) which I shall not review.  
The empirical programme of Pan et al  who include Zeilinger relies on two erroneous presumptions.  In the first place, the joint activation of three different polarization designs $Y_1Y_2X_3, Y_1X_2Y_3,$ and $X_1Y_2Y_3$ cannot be performed on the same triplet of photons, so in deference to the general uncertainty principle, quantum theory explicitly avoids any claims regarding their simultaneous result. It is a ``thought experiment'' to which the implications of locality would pertain, asserting that the value of $Y_2$ for example in a simultaneous instantiation of the $Y_1Y_2X_3$ design must be identical to its value in an $X_1Y_2Y_3$ design activation on the same triple of photons.  This is a claim widely recognized as lying outside the domain of quantum theoretical assertions.  In the second place, in their empirical evaluation of results on three {\it different} triples of photons, the conditions of quantum superposition assure only that the {\it product} values of $Y_1Y_2X_3$ and $X_1Y_2Y_3$ are both identically equal to $-1$, not that any individual multiplicands of the product triples are equal. In particular, the value of $Y_2$ in the first experimental triple need not equal the value of $Y_2$ in the second triple. The implication proposed in their analysis that the value of $X_1X_2X_3$ in any triple run on three distinct triples of photons that meet the GHZ condition in their experiment does not hold.  Their allusion to experimental error as accounting for their mixed results does not wash.  I would be happy to be more explicit in discussion or in another article, but I will leave these comments in this terse form for now.

\end{document}